\documentclass{aa}  
\usepackage{graphicx}
\usepackage{txfonts}
\usepackage{natbib}
\usepackage{color}
\usepackage{hyperref}
\hypersetup{colorlinks=True,citecolor=blue,allcolors=blue}
\usepackage{tabularx}
\usepackage{tabu}
\usepackage{float}
\usepackage{multirow}
\pdfoutput=1 

\begin{document} 
   \title{Probing the structure of the lensed quasar SDSS J1004+4112 through microlensing analysis of spectroscopic data}
\author{C. Fian\inst{1}, J. A. Mu\~noz\inst{1,2}, R. Forés-Toribio\inst{1,2}, E. Mediavilla\inst{3,4}, J. Jim\'enez-Vicente\inst{5,6},  D. Chelouche\inst{7,8}, S. Kaspi\inst{9}, G. T. Richards\inst{10}}
\institute{Departamento de Astronom\'{i}a y Astrof\'{i}sica, Universidad de Valencia, E-46100 Burjassot, Valencia, Spain; \email{carina.fian@uv.es} \and Observatorio Astron\'{o}mico, Universidad de Valencia, E-46980 Paterna, Valencia, Spain  \and Instituto de Astrof\'{\i}sica de Canarias, V\'{\i}a L\'actea S/N, La Laguna 38200, Tenerife, Spain \and Departamento de Astrof\'{\i}sica, Universidad de la Laguna, La Laguna 38200, Tenerife  \and Departamento de F\'{\i}sica Te\'orica y del Cosmos, Universidad de Granada, Campus de Fuentenueva, 18071 Granada, Spain \and Instituto Carlos I de F\'{\i}sica Te\'orica y Computacional, Universidad de Granada, 18071 Granada, Spain \and Department of Physics, Faculty of Natural Sciences, University of Haifa,
Haifa 3498838, Israel \and Haifa Research Center for Theoretical Physics and Astrophysics, University of Haifa,
Haifa 3498838, Israel \and School of Physics and Astronomy and Wise Observatory, Raymond and Beverly Sackler Faculty of Exact Sciences, Tel-Aviv University, Tel-Aviv 6997801, Israel \and Department of Physics, Drexel University, 32 S. 32nd Street, Philadelphia, PA 19104, USA}
 

  \abstract
  {}
   {We aim to reveal the sizes of the continuum and broad emission line (BEL) emitting regions in the gravitationally lensed quasar SDSS J1004+4112 by analyzing the unique signatures of microlensing in this system. Through a comprehensive analysis of 20 spectroscopic observations acquired between 2003 and 2018, we studied the striking deformations of various BEL profiles and determined the sizes of their respective emitting regions.}
   {Our approach involves a detailed analysis of the magnitude differences in the BEL wings and their adjacent continua, and the implementation of a statistical model to quantify the distribution and impact of microlensing magnifications. To ensure a reliable baseline for no microlensing, we used the emission line cores as a reference. We then applied a Bayesian estimate to derive the size lower limits of the Ly$\alpha$, Si IV, C IV, C III], and Mg II emitting regions, as well as the sizes of the underlying continuum-emitting sources.}
   {We analyzed the outstanding microlensing-induced distortions in the line profiles of various BELs in the quasar image A, characterized by a prominent magnification of the blue part and a strong demagnification of the red part. From the statistics of microlensing magnifications and using Bayesian methods, we estimate the lower limit to the overall size of the regions emitting the BELs to be a few light-days across, which is significantly smaller than in typically lensed quasars. The asymmetric deformations in the BELs indicate that the broad-line region is generally not spherically symmetric, and is likely confined to a plane and following the motions of the accretion disk. Additionally, the inferred continuum-emitting region sizes are larger than predictions based on standard thin-disk theory by a factor of $\sim3.6$ on average. The size-wavelength relation is consistent with that of a geometrically thin and optically thick accretion disk.}
   {}


\titlerunning{Dimensions of the accretion disk and BLR in SDSS J1004+4112}
\authorrunning{Fian et al.} 
\maketitle

\section{Introduction} 
\begin{table*}[h]
\centering
\tabcolsep=0.35cm
\caption{Spectroscopic data.}
\begin{tabular}{cccccc} \hline \hline \vspace*{-3mm}\\
Epoch & Date & Image & BEL & Facility & Reference\\ \hline \vspace*{-3mm} \\
1a & 03-05-2003 & A & C IV & ARC 3.5m & \citealt{Richards2004}\\ 
1b & 03-02-2003 & B & C IV& SDSS 2.5m & \citealt{Richards2004} \\
2 & 31-05-2003 & A, B, C , D & Ly$\alpha$, Si IV, C IV, C III], Mg II & Keck I 10m & \citealt{Richards2004} \\
3 & 21-11-2003 & A, B & C IV, C III] & ARC 3.5m & \citealt{Richards2004} \\
4 & 30-11-2003 & A, B & C IV, C III] & ARC 3.5m & \citealt{Richards2004} \\ 
5 & 01-12-2003 & A, B & C IV, C III] & ARC 3.5m & \citealt{Richards2004} \\ 
6 & 22-12-2003 & A, B & C IV, C III] & ARC 3.5m & \citealt{Richards2004} \\ 
7 & 19-01-2004 & A, B & C IV, C III] & WHT 4.2m & \citealt{Gomez2006}\\ 
8 & 26-03-2004 & A, B & C IV, C III] & ARC 3.5m & \citealt{Richards2004b}\\ 
9 & 10-04-2004 & A, B & C IV, C III] & ARC 3.5m & \citealt{Hutsemekers2023}\\ 
10 & 26-04-2004 & A, B & C IV, C III] & ARC 3.5m & \citealt{Hutsemekers2023}\\ 
11 & 13-05-2004 & A, B & C IV, C III] & ARC 3.5m & \citealt{Hutsemekers2023}\\ 
12 & 28-05-2004 & A, B & C IV, C III] & ARC 3.5m & \citealt{Hutsemekers2023}\\ 
13 & 08-12-2004 & A, B & C IV, C III] & ARC 3.5m & \citealt{Hutsemekers2023}\\
14 & 17-12-2004 & A, B & C IV, C III] & ARC 3.5m & \citealt{Hutsemekers2023}\\
15 & 01-05-2006 & A, B & C IV, C III] & ARC 3.5m & \citealt{Hutsemekers2023}\\
16 & 16-05-2007 & A, B, C, D & C IV, C III] & SAO RAS 6m & \citealt{Popovic2020} \\
17 & 12-01-2008 & A, B & Ly$\alpha$, Si IV, C IV, C III] & MMT 6.5m & \citealt{Motta2012}\\
18 & 27-10-2008 & A, B, C, D & C IV, C III] & SAO RAS 6m & \citealt{Popovic2020}\\
19 & 11-03-2016 & A, B & Si IV, C IV, C III] & WHT 4.2m & \citealt{Fian2021blr}\\
20 & 07-02-2018 & A, B, C, D & C IV, C III] & SAO RAS 6m & \citealt{Popovic2020}\\ \hline \vspace*{-5mm}\\
\end{tabular}
\label{data}    
\end{table*}
SDSS J1004+4112 is the first known quasar lensed by a foreground cluster of galaxies and was discovered in the Sloan Digital Sky Survey (SDSS) while searching for large separation lenses (\citealt{Inada2003}). The lensed quasar comprises four bright images (with magnitudes of $i' =18.5$, $18.9$, $19.4$, and $20.1$; see \citealt{Inada2003}) at a source redshift of $z_s = 1.734$ and a maximum separation angle of $14\arcsec.62$ between components B and C, which are produced by a massive cluster at a lens redshift of $z_l = 0.68$. The four lensed images are located near the cluster center, and are well separated from the central galaxy emission. A fifth faint image (component E), located $0\arcsec.2$ from the center of the brightest galaxy in the lensing cluster, was detected in deep HST imaging by \citet{Inada2005} and spectroscopically confirmed by \citet{Inada2008}. \citet{Oguri2010} estimated the magnifications of the images to be $29.7$, $19.6$, $11.6$, $5.8$, and $0.16$ for A, B, C, D, and E, respectively. In addition to the quasar images, multiply imaged background galaxies were identified by \citet{Sharon2005}, \citealt{Oguri2010}, and \citealt{Liesenborgs2009}, and more galaxy members were obtained by \citet{Oguri2004}. The system has been extensively monitored photometrically (\citealt{Fohlmeister2007,Fohlmeister2008,Fian2016,Munoz2022}), and time delays for three of the quasar images (A, B, and C) were measured by \citet{Fohlmeister2007,Fohlmeister2008}. Recently, \citet{Munoz2022} found a time delay of $2458.47 \pm 1.02$ days ($\sim 6.7$ years) between the trailing image D and the leading image C, which is the longest ever measured for a gravitationally lensed quasar. Since its discovery, the lens has been modeled by numerous authors (\citealt{Inada2003,Oguri2004,WilliamsSaha2004,KawanoOguri2006,Fohlmeister2007,Inada2008,Liesenborgs2009,Oguri2010,Mohammed2015,Fores-Toribio2022}), using constraints including the multiple time delays, the position and fluxes of the lensed images, spectroscopy of galaxies in the cluster, and Chandra X-ray observations (\citealt{Ota2006}). In addition, radio, infrared (IR), and ultraviolet (UV) observations have been used to measure the wavelength-dependent flux ratios between the lensed images and to study the cluster and background lensed galaxies (\citealt{Ross2009,Jackson2011,McKean2021,Hartley2021}).

Optical microlensing, caused by stars in the cluster halo or nearby satellites, is known to exist in the lens system \mbox{SDSS J1004+4112}. This phenomenon has been used to determine the size of the continuum-emitting source in the lensed quasar (\citealt{Fohlmeister2008,Fian2016,Foresr2023submitted}). In addition to the differential variation of the brightness of the images in the photometric passbands, microlensing has also been detected in the spectral lines (\citealt{Richards2004, Lamer2006, Gomez-Alvarez2006, Motta2012,  Popovic2020, Fian2018blr, Fian2021blr}). Spectroscopy of the individual components has revealed significant differences in the emission line profiles, with component A showing a strong enhancement in the blue wings of several high-ionization lines relative to the other components. This blue-wing enhancement was interpreted as an indication of microlensing in image A (alternative interpretations such as outflow have been proposed in \citealt{Green2006} and \citealt{Popovic2020}). Recently, \citet{Hutsemekers2023} demonstrated that the deformation of the C IV emission line profile in \mbox{SDSS J1004+4112} can be reproduced by employing simple broad-line region (BLR) models such as a Keplerian disk or an equatorial wind. For this study, we extensively analyzed spectroscopic observations gathered from the literature, significantly expanding the existing research. We incorporated data from five additional observational epochs compared to the most recent study conducted by \citet{Hutsemekers2023}. We broadened the scope to include a wider range of spectral lines instead of only the C IV line, thereby expanding the investigation of the magnitude of microlensing variability in several broad-emission lines (BELs). Additionally, we undertake a comprehensive global analysis that encompasses the line cores, line wings, and the adjacent continuum to enhance our understanding of the diverse light-emitting regions within the quasar.\\

The paper is organized as follows. In Section \ref{2}, we present the spectra collected from the literature. In Section \ref{3}, we describe the data analysis, including measurements of core ratios, continuum variability, and the estimation of variability in the BELs. Section \ref{4} is dedicated to microlensing simulations and the inference of source sizes using Bayesian methods. In Section \ref{5}, we present our main results and discuss them in the context of the dimensions of the continuum and broad-line emitting regions. Finally, our main conclusions are summarized in Section \ref{6}.

\begin{figure*}[p]
\centering
\vspace*{1.4cm}
\includegraphics[width=0.42\textwidth]{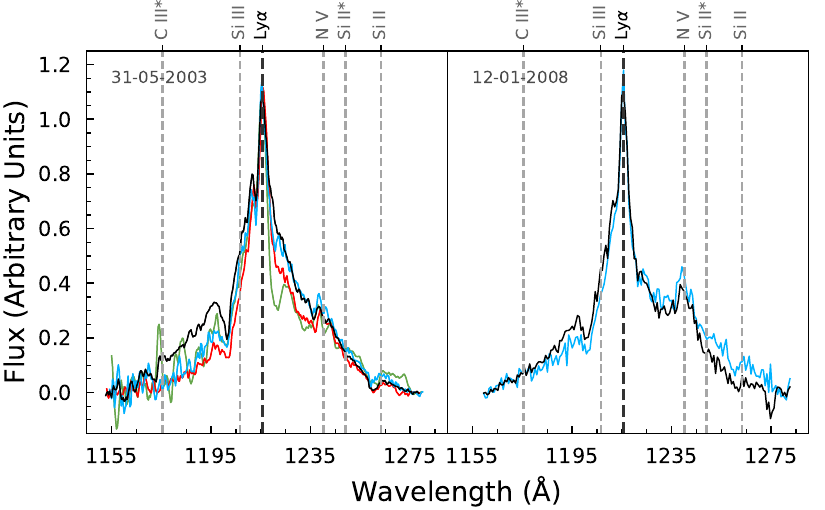}
\includegraphics[width=0.55\textwidth]{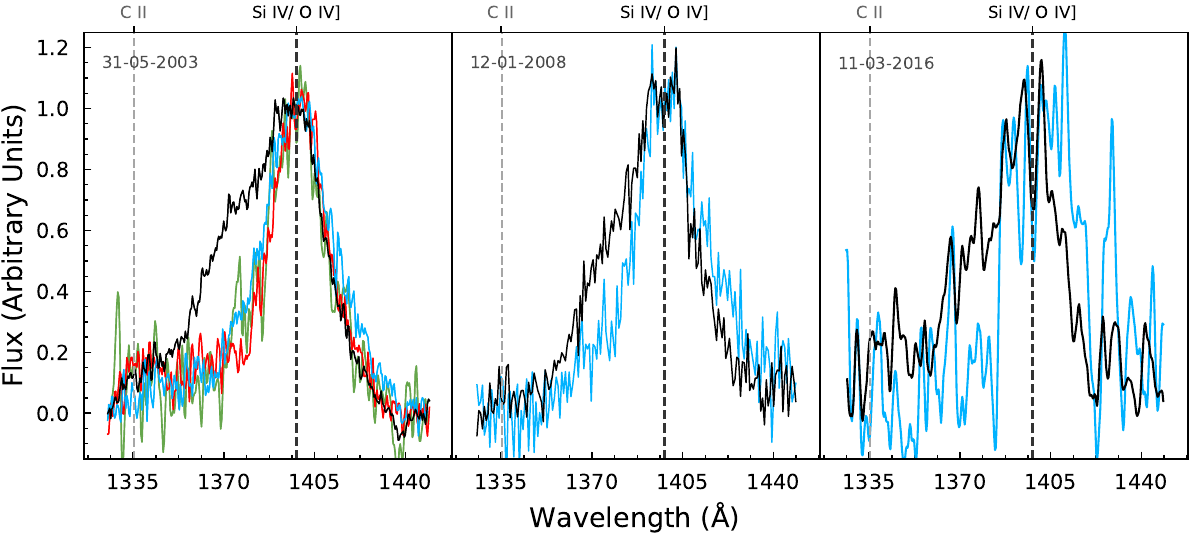}
\includegraphics[width=18.2cm]{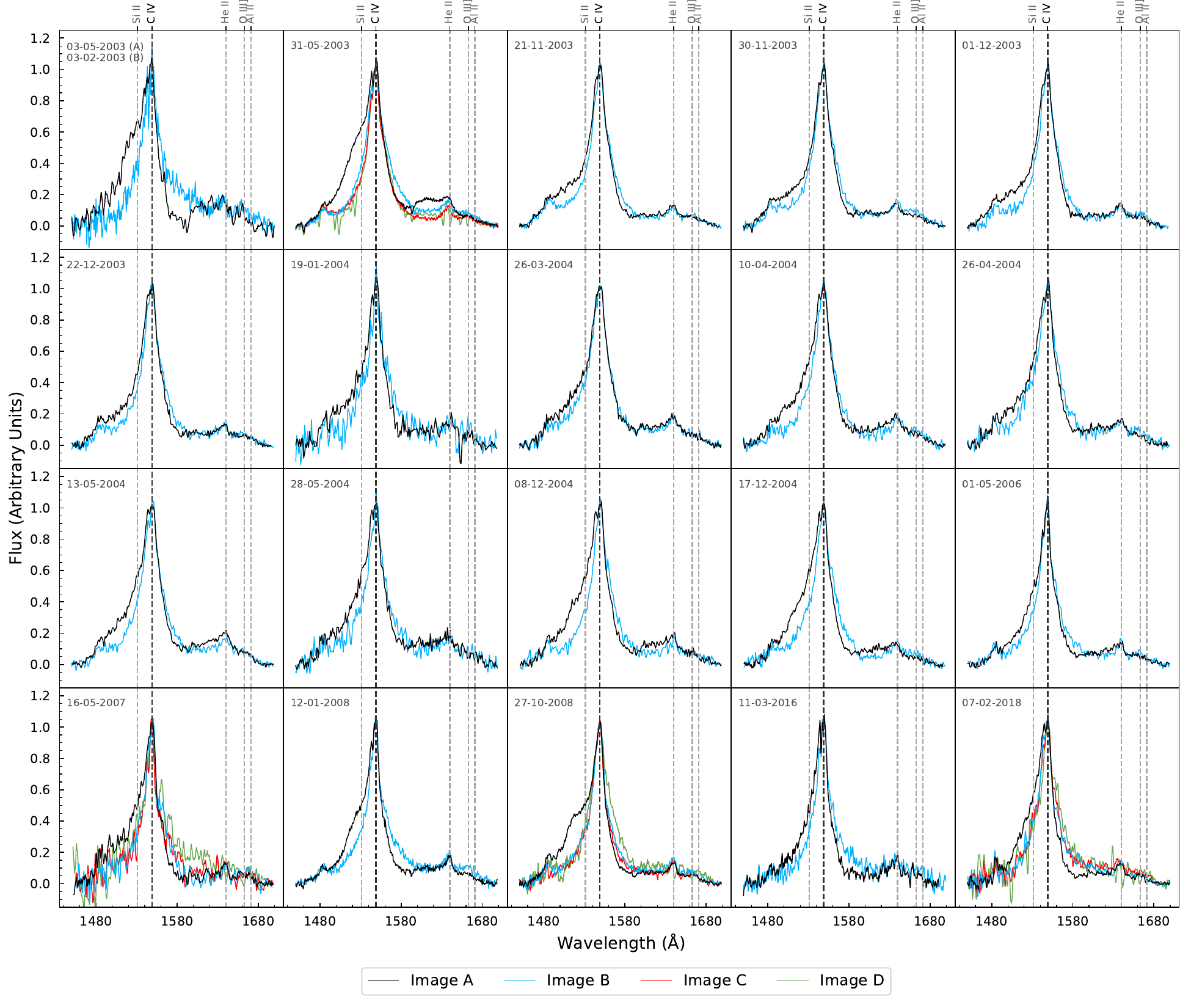}
\caption{Emission line profiles of images A (black), B (blue), C (red), and D (green) at different observational epochs in the rest-frame after the continuum has been subtracted and the line core has been matched (see text). The Ly$\alpha$ line is shown in the top left panel at two different epochs. The top right panel shows the Si IV line in three observational epochs, while the bottom panel displays 20 observational epochs for the C IV line. Observations reveal significant deformations and variability in the blue wings of all emission lines, with image A showing the most pronounced changes. In addition, image A appears to be demagnified in the red wings. The y-axis represents the flux in arbitrary units.}
\label{emissionlines}
\end{figure*}

\begin{figure*}[p]
\centering
\includegraphics[width=18.cm]{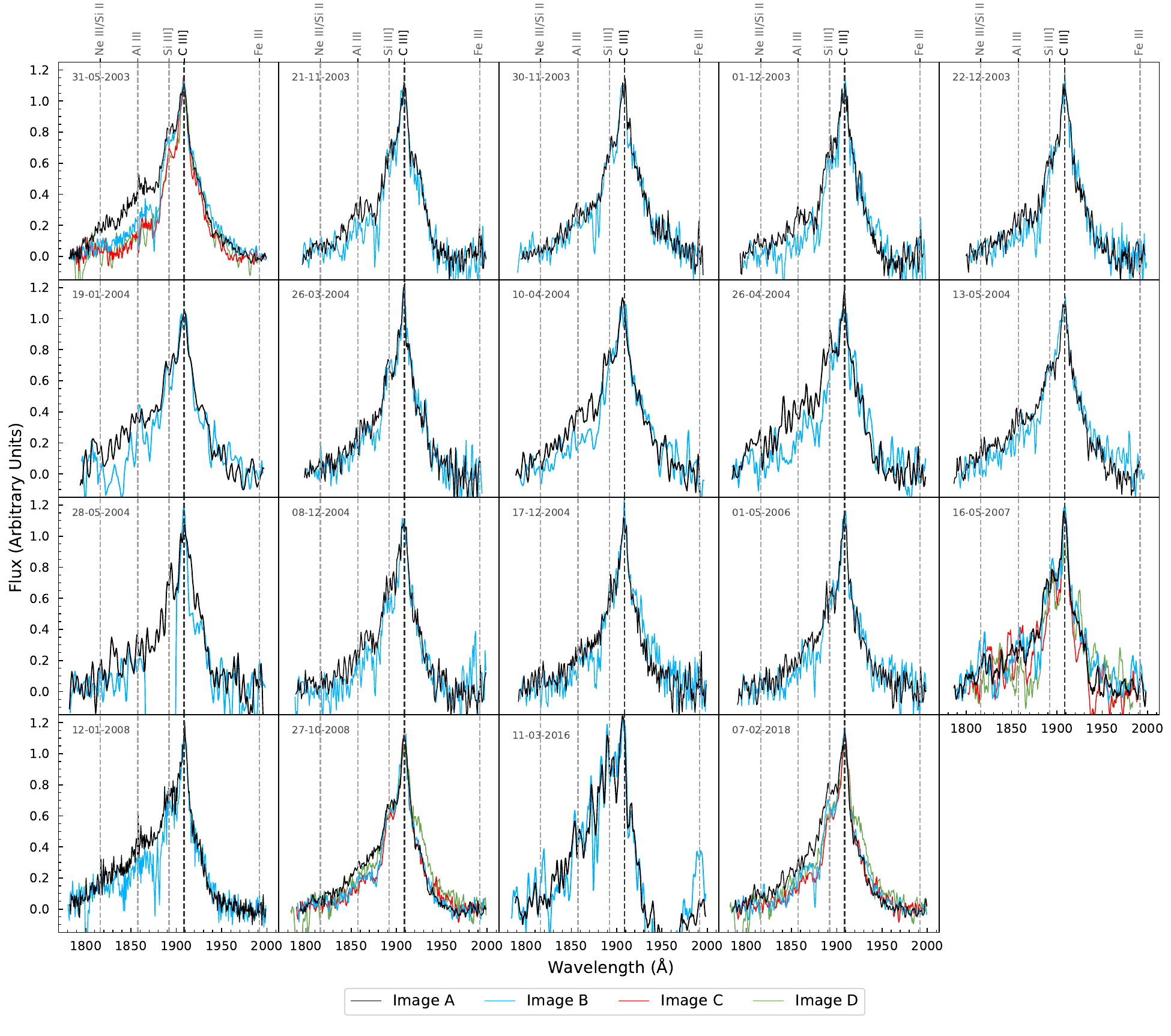}
\caption{Same as Figure \ref{emissionlines}, but displaying 19 observational epochs of the C III] line.}
\label{emissionlines_2}
\includegraphics[width=0.99\textwidth]{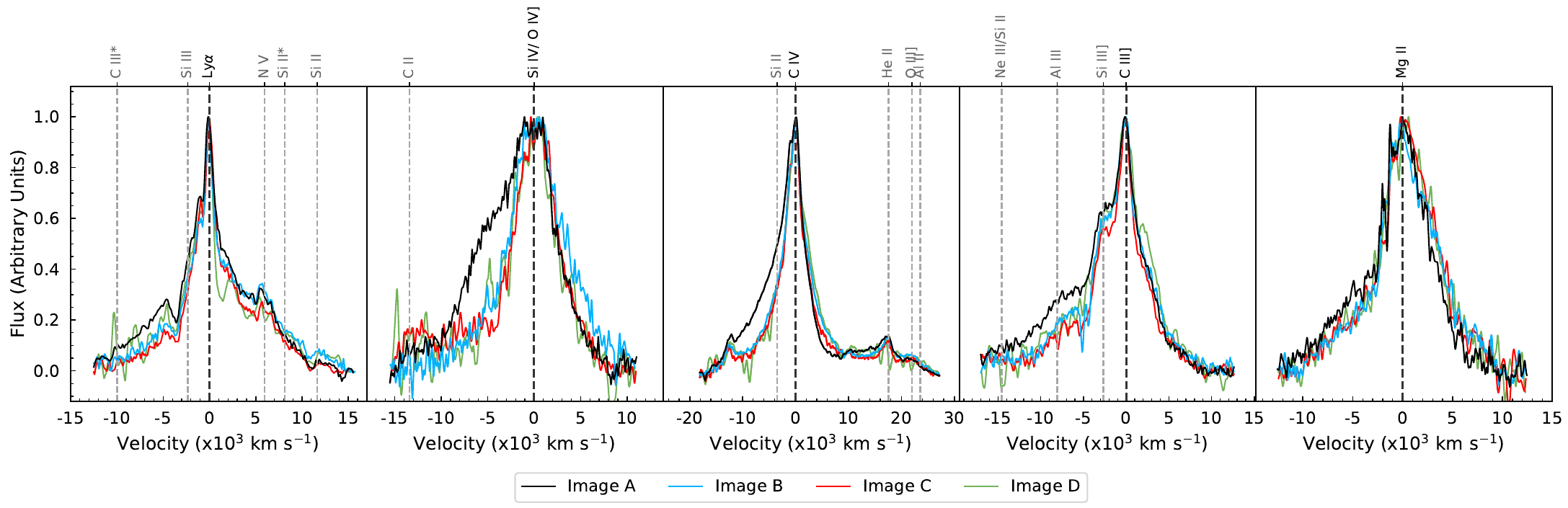}
\caption{Average Ly$\alpha$, Si IV, C IV, C III], and Mg II emission line profiles (from left to right) of the images A (black), B (blue), C (red), and D (green). Observations reveal significant deformations and differences between the images in the blue wings of all emission lines, with image A showing the most pronounced enhancements. In addition, image A appears to be de-magnified in the red wings. The x-axis is represented on a velocity scale, and the y-axis is in arbitrary units of flux. We note that only one epoch of observation (epoch 2) is available for the Mg II line.}
\label{mean_ABCD}
\end{figure*}





\section{Data and observations}\label{2}
We have compiled a collection of rest-frame UV spectra of the images A, B, C, and D of the gravitationally lensed quasar \mbox{SDSS J1004+4112} from the literature. Our dataset includes 20 published spectra, spanning a period of 15 years (from February 2003 to February 2018). These spectra cover a range of typical high- and low-ionization lines found in quasars, including Ly$\alpha$ $\lambda$1216, Si IV $\lambda$1397, C IV $\lambda$1549, C III] $\lambda$1909, and Mg II $\lambda$2798. The data from the literature have already been fully reduced, and information about the observations and references can be found in Table \ref{data}. In Figures \ref{emissionlines}--\ref{emissionlines_2}, we present superpositions of the Ly$\alpha$, Si IV, C IV, and C III] emission line profiles, respectively, corresponding to different images and epochs. In Figure \ref{mean_ABCD}, we show the average of these line profiles for each lensed image, alongside the sole epoch for which we possess reliable data for the Mg II line. Individual epochs with noisy data in the range of the emission lines were excluded before obtaining the average spectra. All wavelengths are given in the quasar rest-frame.\\  

\begin{table*}[h]
	\renewcommand{\arraystretch}{1.2}
	\caption{Line core flux ratios.}
	\begin{tabu} to 1\textwidth {X[c]X[c]X[c]X[c]} 
		\hline
		\hline 
		BEL & A/B & C/B & D/B \\ \hline
		Ly$\alpha$ $\lambda 1216$ & $1.55\pm0.04$ & $0.78^{(*)}$ & $0.27^{(*)}$ \\
		Si IV $\lambda 1397$ & $1.72\pm0.11$ & $0.58^{(*)}$ & $0.22^{(*)}$ \\
		C IV $\lambda 1549$ & $1.56\pm0.12$ & $0.76\pm0.12$ & $0.43\pm0.13$ \\
		C III] $\lambda1909$ & $1.43\pm0.11$ & $0.79\pm0.10$ & $0.56\pm0.20$ \\
		Mg II $\lambda 2798$ & $1.33^{(*)}$ & $0.66^{(*)}$ & $0.38^{(*)}$ \\ \hline
		Mean$^{(**)}$ $\pm \sigma$ & $1.51\pm0.14$ & $0.75\pm0.11$ & $0.44\pm0.17$ \\ \hline
	\end{tabu}\\
	
	\small \textbf{Notes.} $^{(*)}$Only a single epoch of observation is available, resulting in the absence of error bars that represent the variability between multiple observations. $^{(**)}$Weighted arithmetic mean (weight is based on the number of spectroscopic observations for each line).
\label{core_flux_ratios}	
\end{table*}

\begin{table*}[h]
	\renewcommand{\arraystretch}{1.2}
	\caption{Comparison of the estimated core flux ratios with previously reported values from the literature.}
	\begin{tabu} to \textwidth {X[c]X[c]X[c]X[c]X[c]} 
		\hline
		\hline 
		A/B & C/B & D/B & Comment & Reference \\ \hline
		 $1.51\pm0.14$ & $0.75\pm0.11$ & $0.44\pm0.17$ & BEL cores & this work \\
		 $1.40\pm0.02$ & $0.77\pm0.01$ & $0.45\pm0.01$ & infra-red (8 $\mu$m) & \citealt{Ross2009}\\
		 $1.64\pm0.39$ & $0.77\pm0.26$ & $0.85\pm0.27$ & radio (5 GHz) & \citealt{Jackson2020} \\
         $1.60\pm0.15$ & $0.91\pm0.11$ & $0.29\pm0.08$ & radio (5 GHz) & \citealt{Hartley2021} \\
		 $1.36\pm0.47$ & --- & --- & radio (144 MHz) & \citealt{McKean2021}\\ \hline
	\end{tabu}
\label{flux_ratios}	
\end{table*}

Upon visually examining the data, it is immediately apparent from Figures \ref{emissionlines} to \ref{mean_ABCD} that the emission line wings exhibit significant (time-variable) deformations and asymmetrical enhancements. Image A displays the most pronounced differences in comparison to the other images, featuring much stronger blue emission line wings in the high-ionization lines. Image B, on the other hand, appears to have an enhanced red wing as compared to the other images. It is important to note that in Figures \ref{emissionlines} to \ref{mean_ABCD}, we have highlighted the differences in the emission line profiles by subtracting a linear fit of the continuum and normalizing the spectra to the peak of the corresponding emission line. In general, there is a good match between the C III] red wings of different images, indicating that the red part of this line is only weakly affected by either microlensing or intrinsic variability. However, we find changes in the red wing of C IV and/or in the shelf-like feature blueward of He II (at $\sim \lambda$1610) in all the images, which appear with different intensities in different epochs. The origin of this feature is uncertain, with possible explanations including an extreme C IV red wing or He II blue wing, or an as-yet-unidentified species (e.g., \citealt{Fine2010} and references therein). In the case of images A and B (with a time delay of $44$ days), the observed anomalies are likely caused by microlensing, whereas in the case of images C and D (with a time delay of $-2458$ days), the differences may be explained by intrinsic variability combined with the large time delay between the images, and a possible contribution from microlensing.

\section{Data analysis}\label{3}
\subsection{Core ratio measurements}\label{calculate-core}
Emission line cores originate from large, spatially extended regions that are not significantly affected by microlensing and intrinsic variability (\citealt{Guerras2013,Fian2018blr}). Measuring the core flux ratios between lensed images is an effective way to establish a baseline that is free of microlensing effects, which is crucial for determining the microlensing-based sizes of different emitting regions in quasars. In this study, we focus on the high-ionization lines Ly$\alpha$, Si IV, and C IV, and the low-ionization lines C III] and Mg II. The continuum for each image and each emission line is removed by fitting a straight line to the continuum on both sides of the emission line and subtracting it from the spectrum. To account for the varying widths of the emission lines, we use windows of varying widths for the continuum estimate for each line and each quasar image, avoiding regions of known emission features. The core fluxes are then defined by a narrow interval ($8\AA$ for Ly$\alpha$, $10\AA$ for Si IV and C IV, $12\AA$ for C III], and $20\AA$ for Mg II) centered on the peak of the line. In Table \ref{core_flux_ratios}, we list the average core flux ratios ($\pm1\sigma$ standard deviation) between images A and B, C and B, and D and B. Although extinction may affect the emission lines, we found no discernible wavelength dependence in the core flux ratios, which suggests that extinction is unlikely to have played a significant role in our estimates. Our inferred core flux ratios match well with IR flux ratios obtained by \citet{Ross2009} and are consistent (within uncertainties) with radio flux ratios observed by \citet{Jackson2020}, \citet{Hartley2021}, and \citet{McKean2021} (see Table \ref{flux_ratios}).

\begin{table*}[h]
	\renewcommand{\arraystretch}{1.2}
	\caption{Differential microlensing measurements in the continuum.}
	\begin{tabu} to \textwidth {X[c]X[c]X[c]X[c]X[c]} 
		\hline
		\hline 
		$\lambda_{cont}$ (\AA) & $\Delta \lambda_{cont}$ (\AA) & $\Delta \mu_{AB}$ (mag) & $\Delta \mu_{CB}$ (mag) & $\Delta \mu_{DB}$ (mag) \\  
		(1) & (2) & (3) & (4) & (5) \\ \hline
		$1157$ & $10$ & $+0.34\pm0.45$ & $+0.43$ & $+0.21$ \\
		$1281$ & $7$ & $+0.24\pm0.34$ & $+0.54$ & $+0.67$\\
		$1325$ & $8$ & $+0.53\pm0.28$ & $+0.33$ & $+0.11$ \\
        $1450$ & $10$ & $+0.23\pm0.25$ & $+0.45\pm0.24$ & $+0.01\pm0.29$ \\ 
		$1702$ & $13$ & $+0.20\pm0.21$ & $+0.41\pm0.33$ & $-0.30\pm0.40$ \\
		$1784$ & $18$ & $+0.07\pm0.21$ & $+0.37\pm0.23$ & $-0.08\pm0.28$ \\
		$1989$ & $17$ & $+0.07\pm0.19$ & $+0.37\pm0.20$ & $-0.15\pm0.33$ \\
		$2672$ & $24$ & $-0.07\pm0.17$ & $+0.29$ & $+0.20$ \\ 
		$2915$ & $26$ & $-0.06\pm0.15$ & $+0.26$ & $+0.10$ \\ \hline 
	\end{tabu}\\
		
		\small \textbf{Notes.} Col. (1): Average central wavelength of the continuum. Col. (2): Average wavelength window used for computing the magnitude differences in the continuum. Cols. (3)--(5): Average $\pm$ 1$\sigma$ differential microlensing measurements between the images A and B, C and B, and D and B, respectively.
\label{microlensing_continuum}	
\end{table*}

\begin{table*}[h]
	\renewcommand{\arraystretch}{1.2}
	\caption{Differential microlensing measurements in the BEL wings.}
	\begin{tabu} to \textwidth {X[c]X[c]X[c]X[c]X[c]X[c]} 
		\hline
		\hline 
		BEL & Feature & Window (\AA) & $\Delta \mu_{AB}$ (mag) & $\Delta \mu_{CB}$ (mag) & $\Delta \mu_{DB}$ (mag) \\  
		(1) & (2) & (3) & (4) & (5) & (6) \\ \hline
		Ly$\alpha$ $\lambda 1216$ & blue wing & 28 & $-0.44\pm0.17$ & $+0.29$ & $-0.27$ \\
		 & red wing & 28 & $+0.13\pm0.08$ & $+0.20$ & $+0.23$ \\ \hline 
		Si IV $\lambda 1397$ & blue wing & 33 & $-0.80\pm0.12$ & $+0.18$ & $-0.11$ \\
		 & red wing & 33 & $+0.48\pm0.12$ & $+0.37$ & $+0.17$ \\ \hline
		C IV $\lambda 1549$ & blue wing & 36 & $-0.53\pm0.16$ & $+0.06\pm0.08$ & $+0.00\pm0.31$ \\
		 & red wing & 36 & $+0.48\pm0.27$ & $+0.13\pm0.27$ & $-0.22\pm0.46$ \\ \hline
		C III] $\lambda1909$& blue wing & 45 & $-0.37\pm0.13$ & $+0.13\pm0.19$ &$-0.06\pm0.14$\\
		 & red wing & 45 & $+0.19\pm0.15$ & $+0.00\pm0.26$ & $-0.33\pm0.32$ \\ \hline
		Mg II $\lambda 2798$ & blue wing & 65 & $-0.37$ & $-0.03$ & $-0.12$ \\
		 & red wing & 65 & $+0.45$ & $-0.02$ & $+0.08$ \\ \hline 
	\end{tabu}\\
		
		\small \textbf{Notes.} Cols. (1)--(2): Emission line and line wing. Col. (3): Wavelength window of the line wing. Cols. (4)--(6): Average $\pm$ 1$\sigma$ differential microlensing between the images A and B, C and B, and D and B, respectively.
\label{microlensing_BEL}	
\end{table*}

\subsection{Continuum variability measurements}\label{calculate-continuum}
Microlensing, which is sensitive to the size of the source region (with smaller regions showing larger magnifications), can provide important constraints on the structure and kinematics of quasar accretion disks. To quantify the effect of microlensing on the continuum, it is necessary to separate it from the effects of macro-lensing magnification (due to the smooth lensing potential) and extinction. Each lensed quasar image has the same intrinsic spectrum but experiences different extinction as light associated with each image follows a different path through the lens galaxy, encountering varying amounts of dust and gas (\citealt{Falco1999,Motta2002,Munoz2004,Munoz2011}). The macro-magnification produced by the lens galaxy and the differential extinction between the components of the lensed quasar are independent of the source size and affect not only the continuum flux ratios but also the emission line fluxes (\citealt{Motta2002,Mediavilla2005,Mediavilla2009,Mediavilla2011}). We attempt to correct for these effects by estimating the offsets between the continuum adjacent to the emission lines, $(m_x-m_B)_{cont}$, and the magnitude differences of the emission line cores, $(m_x-m_B)_{core}$, between the images $x$ (where $x = A, C, D$) and image B, \mbox{$\Delta \mu_{cont} = (m_x-m_B)_{cont}-(m_x-m_B)_{core}$}. The cores of the emission lines are produced by material spread over a wide region (narrow-line region and outer regions of the BLR) which is typically large enough to be insensitive to microlensing by solar mass objects and can be used as a baseline for no microlensing (\citealt{Guerras2013,Fian2018blr}). Recently, \citet{Hutsemekers2023} reported a small-to-negligible de-magnification of the core of the C IV line in the lensed system SDSS J1004+4112. This finding further supports the reliability of using the BEL cores as a reference for no microlensing. We selected image B as the reference image as it is less affected by microlensing variability than image A (see \citealt{Hutsemekers2023}), and also has a larger number of observation epochs compared to images C and D. Since the wavelength differences between the line cores and the chosen wavelength intervals for the continuum (see Table \ref{microlensing_continuum}) are relatively small, this estimator certainly removes the effects of the macro-magnification and extinction (see, e.g.,  \citealt{Guerras2013}). Apart from these effects, other phenomena such as intrinsic variability and contamination by the lens galaxy can also produce chromatic variations in the flux of lensed quasars, mimicking microlensing. In the case of the lensing cluster \mbox{SDSS J1004+4112}, the components A, B, C, and D pass far from the main galaxy cluster members and contamination from the continuum of these galaxies is negligible. The magnitude differences between images C--B and D--B may be contaminated by intrinsic variability modulated by the long lens time delays between those image pairs, making it difficult to distinguish between extrinsic and intrinsic variations. When estimating the size of the continuum-emitting region, these (wavelength-dependent) changes should, if possible, be avoided or at least estimated. In Table \ref{microlensing_continuum}, we list the central values used to fit the continuum together with the average magnitude difference ($\pm1\sigma$ standard deviation) at that wavelength.

		

\subsection{BLR variability measurements}\label{calculate-BLR} 
\begin{figure*}[h!]
\includegraphics[width=0.98\textwidth]{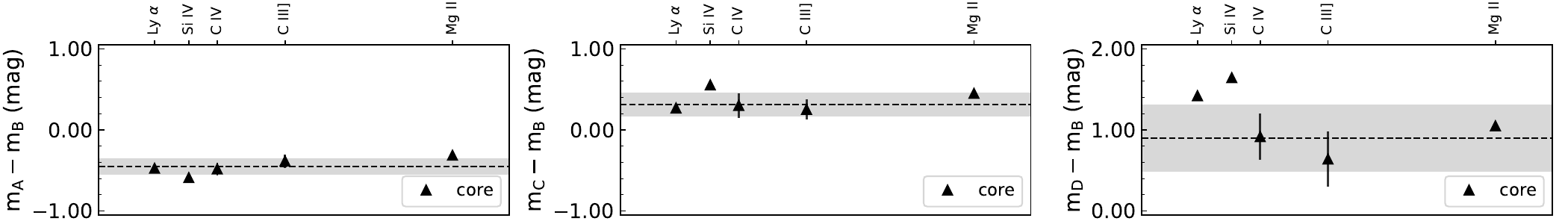}
\includegraphics[width=0.98\textwidth]{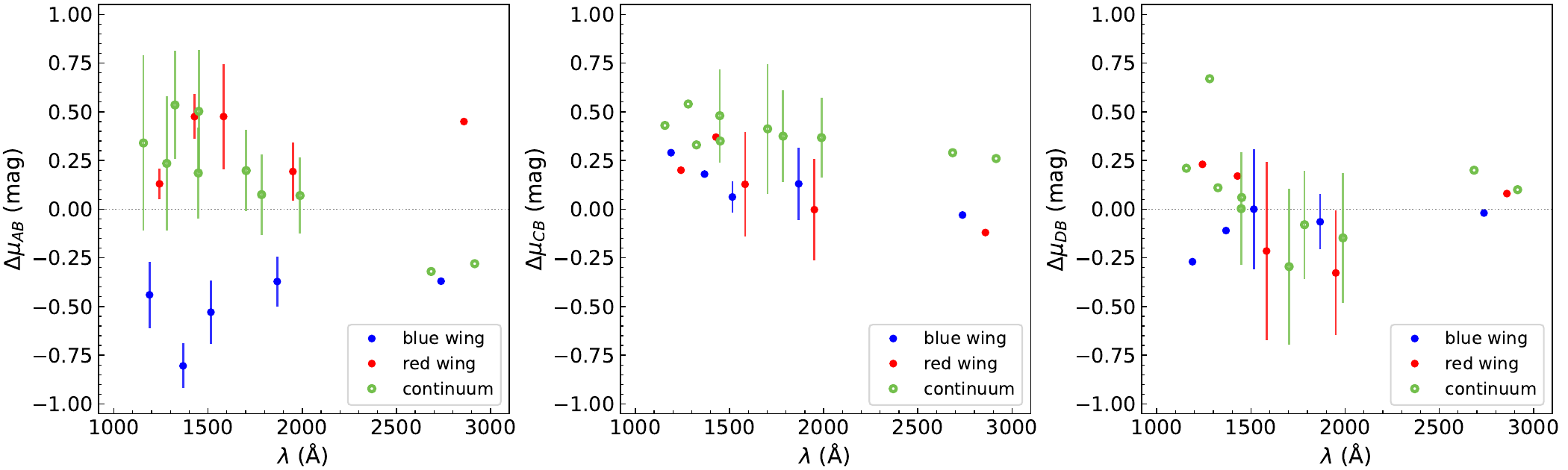}
\caption{Average magnitude differences in the line cores (upper panels), and differential microlensing estimates in the continuum and emission line wings (lower panels) for the image pairs A--B (left), C--B (middle), and D--B (right). Calculations are detailed in Sections \ref{calculate-core}--\ref{calculate-BLR}, and the corresponding results are presented in Tables \ref{core_flux_ratios}, \ref{microlensing_continuum}, and \ref{microlensing_BEL}. Black triangles show the magnitude differences in the line cores, while the black horizontal lines represent the average magnitude differences. The gray-shaded intervals denote the standard deviation. Notably, no discernible wavelength trend is observed in the magnitude differences of line cores among all image pairs. The green-colored data points represent the differential microlensing estimates in the continuum at various wavelengths for the A--B, C--B, and D--B image pairs, respectively. One intriguing observation is the substantial offset of around 0.9 mag in the A--B microlensing measurements between the blue and red wings (shown as blue- and red-colored data points, respectively) of multiple emission lines. This offset suggests that image A experiences magnification in the blue wing and de-magnification in the red wing.} 
\label{cores}
\end{figure*}
To estimate the minimal dimension of the BEL emitting regions, we normalize the continuum-subtracted spectra for all images and all epochs to match the core of the emission line defined by the flux within a narrow interval centered on the peak of the line. Under the assumption that the line cores can be used as a reference that is little affected by microlensing and intrinsic variability, the comparison of the line wing fluxes between pairs of images at the same epoch allows for an estimation of the size lower limit of the emitting region. We estimate the microlensing in the line wings ($\Delta \mu_{wing} = (m_x-m_B)_{wing}-(m_x-m_B)_{core}$, where $x = A,C,D$) on either side of the emission line peak, corresponding to a velocity range of \mbox{$3000-10000$ km s$^{-1}$} (see Table \ref{microlensing_BEL}). Completely distinguishing microlensing from intrinsic variability is not possible without observations separated by the exact time delay between images. As a result, intrinsic variability, in combination with the substantial time delays between the image pairs C--B ($\sim 2$ years) and D--B ($\sim 4.5$ years), may mimic microlensing. However, we assume that the effect of intrinsic variability on the magnitude differences observed between images A and B is negligible since the time delay between this image pair is small ($\sim 40$ days), making it plausible that most of the observed \mbox{A--B} magnitude differences in the continuum and BEL wings are caused by microlensing.



We observe significant changes in the blue wings of Si IV, C IV, and C III] in image A, as well as in the red wing of Ly$\alpha$ and the shelf-like feature at $\sim\lambda1610$ (blueward of He II). Image B appears to vary less, except in the red wing of C IV and both wings of C III]. It is worth noting that the results regarding C III] should be interpreted with caution as (i) the S/N of this emission line is lower than for the high-ionization lines studied in this work, (ii) the blue wing of C III] is blended by Si III], Al III, and Ne III/Si II, (iii) the extreme blue wing of C III] is contaminated by Fe III, and (iv) the presence of (highly variable) \mbox{Fe II} and \mbox{Fe III} lines might influence the continuum adjacent to this line. In the case of images C and D, we detect variability in the (extreme) red wing of C IV, whereas the blue wing of C IV and both wings of C III] do not show significant variability.  

In Figure \ref{cores}, the continuum, the emission line wings, and the emission line core magnitude differences between the A--B, \mbox{C--B}, and D--B image pairs are presented as a function of wavelength. The magnitude differences in the line cores of all image pairs show no significant trend with wavelength and are distributed around $\langle A-B \rangle = -0.46 \pm 0.10 $ mag, $\langle C-B \rangle = 0.32 \pm 0.15 $ mag, and $\langle D-B \rangle = 0.90 \pm 0.41 $ mag. This supports the assumption that the line cores are relatively insensitive to microlensing, intrinsic variability, and extinction, thereby serving as a reliable baseline for no microlensing magnification. The global offset between the adjacent continua on either side of the emission lines and the line cores of the A--B image pair can be attributed to microlensing. The A--B magnitude differences corresponding to the continuum and the red wing (from Si IV toward C III]) show a decreasing trend with wavelength, indicating evidence of chromatic microlensing. Observations reveal an opposite trend at wavelengths shorter than Si IV, which could be potentially attributed to the substantial influence of the Rayleigh scattering wings of the Ly$\alpha$ line contributing to the signal. The modest amplitudes of microlensing observed in the A--B continuum, which were previously viewed as a challenge for interpreting the blue wing enhancements in terms of microlensing (\citealt{Gomez2006}), are now supported by additional evidence, such as the lack of dust extinction, the presence of chromatic magnification changes in the A--B continuum, and the variability detected in the emission line wings. These findings lend strong support to the hypothesis that microlensing is responsible for the enhancements seen in the \mbox{A--B} blue wings. The C--B magnitude differences corresponding to the emission line wings also depict a decreasing trend with wavelength, while neither the C--B magnitude differences in the continuum nor the D--B magnitude differences in the continuum and BEL wings show chromatic changes. However, it is worth noting that intrinsic variability combined with the large time delay between those image pairs could affect the observed magnitude differences, making it difficult to interpret the results. An interesting finding is the large offset ($\sim 0.9$ mag) between the A--B magnitude differences in the blue and red wings of several emission lines, while for the C--B and D--B image pairs the observed offset is relatively small ($\sim 0.1$ mag and $\sim 0.3$ mag, respectively). 

\section{Microlensing simulations}\label{4}
\subsection{Magnification maps}
To simulate the microlensing of extended sources, we utilized the Fast Multipole Method -- Inverse Polygon Mapping\footnote{\href{https://gloton.ugr.es/microlensing/}{https://gloton.ugr.es/microlensing/}} (FMM--IPM) algorithm described in \citet{Jimenez2022} to create microlensing maps for images A and B. This novel technique combines the FMM algorithm of \citet{Greengard1987} for ray deflection calculations with the IPM algorithm of \citet{Mediavilla2006,Mediavilla2011ipm} for the calculation of the magnification map. Our simulations are based on $3000\times3000$ pixel$^2$ maps, spanning $100\times100$ Einstein radii$^2$ on the source plane. The value of the Einstein radius for this system is $R_E = 2.35\times 10^{16} \sqrt{M/0.3 M_\odot}\ \mathrm{cm} = 9.1 \sqrt{M/0.3 M_\odot}$ lt-days at the lens plane (see \citealt{Mosquera2011}). The maps have a resolution of $0.3$ lt-days per pixel, which effectively samples the optical accretion disk of the quasar. The magnification maps for each quasar image are characterized by the local shear $\gamma$ and the local convergence $\kappa$, with the latter being proportional to the surface mass density. The local convergence can be broken down into two components: $\kappa = \kappa_c + \kappa_\star$, where $\kappa_c$ represents the convergence due to continuously distributed matter (e.g., dark matter) and $\kappa_\star$ represents the convergence due to stellar-mass point lenses (e.g., microlens stars in the galaxy). The values of $\kappa$ and $\gamma$ for images A and B were obtained from the study of \citet{Fores-Toribio2022} and are presented in Table \ref{macromodel} for reference.\\ 
\begin{table}[h]
	\renewcommand{\arraystretch}{1.2}
	\caption{Macro-model parameters.}
	\begin{tabu} to 0.49\textwidth {X[c]X[c]X[c]X[c]} 
		\hline
		\hline 
		Image & $\kappa$ & $\gamma$ & $\alpha$ \\ 
		(1) & (2) & (3) & (4) \\ \hline
		A & $0.729$ & $0.333$ & $0.05$ \\
		B & $0.651$ & $0.233$ & $0.05$ \\ \hline
	\end{tabu}\\
	
	\small \textbf{Notes.} Col. (1): Lensed quasar image. Cols. (2)--(4) Convergence $\kappa$, shear $\gamma$, and fraction of mass in stars $\alpha$ at the quasar image positions. The values of $\kappa$, $\gamma$, and $\alpha$ were taken from \citet{Fores-Toribio2022} and from For\'es-Toribio et al. in preparation.
\label{macromodel}	
\end{table}
\begin{figure*}
\includegraphics[width=0.98\textwidth]{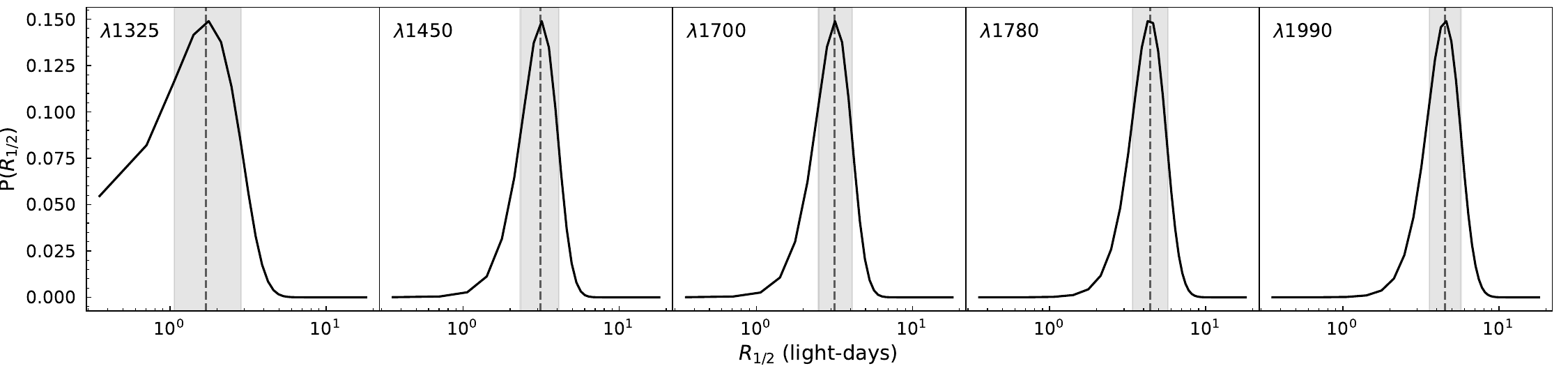}
\caption{PDFs of the half-light radius ($R_{1/2}$) emitting the continuum at different wavelengths. The vertical dashed lines indicate the expected size of the emission region using a uniform logarithmic prior, while the gray-shaded regions represent the one-sigma intervals. A visible trend with wavelength is apparent, with larger sizes observed at longer wavelengths.}
\label{continuum}
\end{figure*}
The value of $\alpha\equiv \kappa_\star/\kappa$, also known as the fraction of mass in compact objects, is a measure of the relative contribution of stars to the total mass in the lens galaxy. In this study, we used a relatively small value of $\alpha=0.05$ for the surface mass density of stars. This value was chosen as the lensed images are located in peripheral regions that are farther away from the center of the lensing galaxy cluster. In these regions, the lensing effect is primarily determined by the distribution of dark matter and hot gas, rather than stars (see also \citealt{Foresr2023submitted}). We randomly distributed stars of a mass of $M = 0.3 M_\odot$ across the microlensing patterns to create a convergence of $5\%$. 

\subsection{Source profile}
To model the structure of the unresolved quasar, we use circular Gaussian profiles ($I(R)\propto \exp (-{R}^{2}/2r_s^{2})$) to represent the luminosity of the emitting regions. The magnifications experienced by a finite source of size $r_s$ are then found by convolving the magnification maps with Gaussian profiles with sigma of $r_s$. It is widely believed that the specific shape of the source's emission profile is not important for microlensing flux variability studies, as the results are mainly controlled by the half-light radius rather than by the detailed intensity profile (\citealt{Mortonson2005}). For Gaussian profiles, the characteristic size $r_s$ is related to the half-light radius by $R_{1/2} = 1.18 r_s$. We convolve the maps with Gaussians of 18 different sizes, logarithmically spanning an interval between 0.3 and 15 lt-days for a mean stellar mass $\langle M\rangle = 0.3 {M}_{\odot}$. As lengths are measured in Einstein radii, all estimated sizes can be rescaled accordingly for a different mean stellar mass using $r_s \propto \sqrt{\langle M\rangle}$. The displacement of an extended source across the magnification patterns is equivalent to a point source moving across a map that has been smoothed by convolution with the source's intensity profile. Strong microlensing anomalies are indication of a compact source, whereas low magnifications could be due to a large source size or due to the location of the source in a relatively calm region of the magnification map. After convolution, we normalize each magnification map by its mean value, and histograms of the normalized maps represent the expected microlensing variability. Finally, by cross-correlating the histograms of image B from the histograms of image A (see \citealt{Fian2016}), we construct the microlensing difference histograms A--B for different values of $r_s$. These simulated microlensing difference histograms can be compared with the experimental values as described in Section \ref{3}.

\subsection{Bayesian source size estimation}\label{statistics}
Given the estimates of differential microlensing in the wings and adjacent continua of different emission lines between lensed images, we can infer the size of their emission region. To accomplish this, a statistical method was utilized where each microlensing measurement was treated as a single epoch event. We then use all available epochs of observation to compute the joint microlensing probability, $P(r_s)$, and obtain an average estimate of the size, following the procedures outlined in \citet{Guerras2013} and \citet{Fian2018blr,Fian2021blr}. It is important to mention that the separation of the line emission into two parts is consistent with the hypothesis that the BLR comprises a flat inner region giving rise to the line wings, surrounded by a larger three-dimensional structure that produces the line core (refer, for instance, to \citealt{Popovic2004}). As such, microlensing-based size measurements for the region emitting the line wings should be considered as approximate lower limits rather than an exact size measurement of the BEL region. 

\section{Results and discussion}\label{5}
Taking into account that microlensing is sensitive to the size of the source, we will utilize our determinations of microlensing magnification amplitudes to estimate the size of the continuum-emitting region at different wavelengths, as well as the minimal size of the emission region for various BELs in the \mbox{SDSS J1004+4112} lensed quasar. The process of inferring differential microlensing from the analysis of spectroscopic data in lensed quasars can be challenging due to the presence of intrinsic variability and the fact that intrinsic flux variations are time-delayed between images. The deformations of the BELs caused by intrinsic variability could mimic microlensing, thereby leading to inaccurate measurements of the source sizes. In the case of the lensed quasar SDSS J1004+4112, this issue is particularly pronounced for images C and D, which have long time delays ($\sim 6.7$ years) compared to the other images in the system. To avoid misinterpretation of the measured signal, we will only use the magnitude differences between images A and B, as they have a short time lag ($\sim 44$ days) and the differences can likely be attributed to microlensing. Additionally, we will only consider observations that are separated in time by more than the source crossing time, which is approximately three months (see \citealt{Mosquera2011}) in the SDSS J1004+4112 lens system due to the high effective transverse velocity of the source.
\subsection{Continuum-emitting region size}\label{calcont}
Utilizing the estimates of differential microlensing between images A and B in the continuum adjacent to the BELs, we are able to constrain the size of the continuum-emitting region at different wavelengths. By applying Bayesian methods, outlined in Section \ref{statistics}, we use a uniform logarithmic prior on $r_s$ to estimate the probability of $r_s$ given the measured microlensing magnification. The resulting probability density functions (PDFs) can be observed in Figure \ref{continuum}. These distributions allow us to determine the 68\% confidence size estimates for the continuum sources at various wavelengths. The sizes of the continuum-emitting regions are summarized in Table \ref{continuumsize}.

\begin{table}[h]
	\renewcommand{\arraystretch}{1.4}
	\caption{Half-light radius $R_{1/2}$ of the continuum-emitting region.}
	\begin{tabu} to 0.49\textwidth {X[c]X[c]X[c]X[c]} 
		\hline
		\hline 
		$\lambda_{cont}$  & $R_{1/2}$ & $R_{1/2}^{\,th}$ & $f$ \\
        ($\AA$) & (lt-days) & (lt-days) & \\ 
		(1) & (2) & (3) & (4) \\ \hline
		1325 & $1.7_{-0.6}^{+1.2}$ & $0.69$ & $2.5$ \\
		1450 & $3.1_{-0.8}^{+1.0}$ & $0.78$ & $4.0$ \\ 
        1700 & $3.2_{-0.7}^{+0.9}$ & $0.96$ & $3.3$ \\ 
        1780 & $4.4_{-1.0}^{+1.3}$ & $1.02$ & $4.3$ \\ 
        1990 & $4.5_{-0.9}^{+1.2}$ & $1.18$ & $3.8$ \\ \hline
	\end{tabu}\\
	
	\small \textbf{Notes.} Col. (1): Continuum wavelength in the quasar's rest-frame. \mbox{Col. (2):} Half-light radius values obtained using a logarithmic prior on the size, expressed in units of $\sqrt{M/0.3 M_\odot}$ lt-days. Col. (3): Theoretical accretion disk size converted to half-light radii. Col. (4): Overestimation factor of microlensing-based disk sizes, $f=R_{1/2}/R_{1/2}^{\,th}$.
\label{continuumsize}	
\end{table}
\begin{figure*}[h!]
\includegraphics[width=0.98\textwidth]{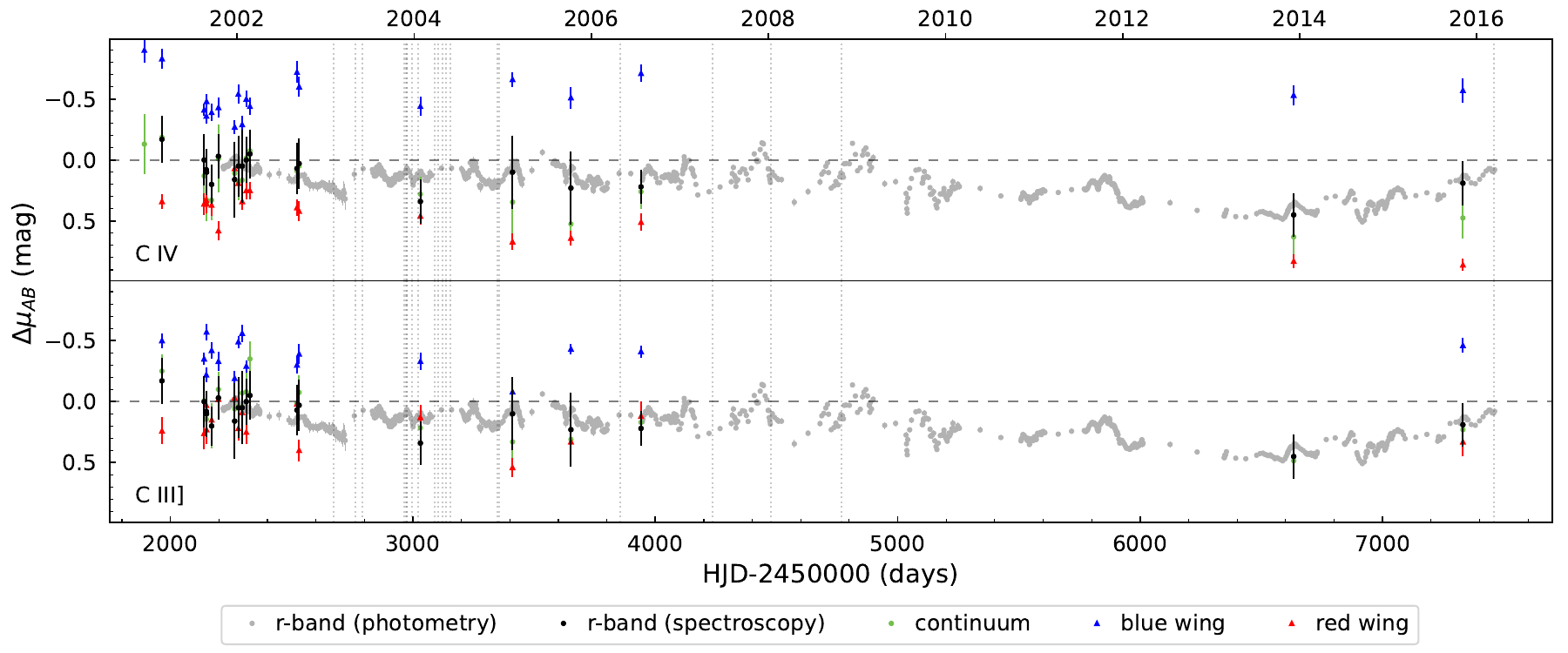}
\caption{Comparison of the microlensing light curves. The optical r-band light curves are shown for the photometric data (gray) and the spectroscopic data (black). The continuum adjacent to the emission line is displayed in green color, and the BEL wings of C IV (top) and C III] (bottom) in blue and red colors, respectively. We want to emphasize the remarkable consistency between the microlensing light curves obtained from spectroscopic data for the r-band emitting region and the results obtained from photometric monitoring data. We note that all observations are presented in the time frame of the leading image C.}
\label{comparison_rband}
\end{figure*}
To validate the accuracy of our results, we have compared the size of the region emitting the r-band continuum inferred from the spectroscopic data presented in this work with the size derived using 14.5 years of photometric monitoring data (see \citealt{Munoz2022}). The results obtained from the two different datasets are in good agreement with each other, with $R_{1/2} = 7.1_{-3.7}^{+7.4}$ lt-days for the spectroscopic data and $R_{1/2} = 5.3_{-0.7}^{+1.3}$ lt-days for the photometric data (see also \citealt{Foresr2023submitted}). We would like to highlight the exceptional overlap of the microlensing light curves of the r-band emitting region obtained from spectroscopic data with the photometric monitoring data, as demonstrated in Figure \ref{comparison_rband}. 
\subsubsection{Theoretical disk size}
In current standard models, the accretion disk is considered to be a geometrically thin and optically thick disk (\citealt{Shakura1973}), radiating thermally with a temperature profile of $T\sim R^{-3/4}$ (\citealt{Shields1978}). This scenario predicts that the hotter, UV-emitting region ($\leq 3000$\AA) is located closer to the center, while the cooler, optically and near-IR emitting regions ($\sim3000-10000$\AA) are located farther out. Variations in the energetic, short-wavelength emission from the X-ray emitting corona and the inner edge of the disk can irradiate the outer annuli and drive variations at longer wavelengths, delayed by the light travel time across the disk (e.g., \citealt{Krolik1991}). For a temperature profile of a standard disk (since the delay $\tau \sim R/c$ and, from Wien's law, $\lambda \propto 1/T$), the disk sizes scale as $R_{\lambda} \propto \lambda^{\ \beta}$, where $\beta=4/3$. This model provides definite predictions about the theoretical time lags between short-wavelength and long-wavelength variations according to a given temperature-radius relation, based on the object's SMBH mass and mass accretion rate. We compare the microlensing-based size estimates of the continuum-emitting regions with model predictions for thermal reprocessing following the method described by \citet{Fausnaugh2016} and \citet{Edelson2017}. Since SMBH mass estimates for lensed quasars are highly uncertain, we substitute the product of SMBH mass and mass accretion rate with the target's optical luminosity $L_{opt}$ (see Eq. (7) in \citealt{Davis2011}; for a detailed derivation see \citealt{Fian2022}). This allows us to use the Shakura-Sunyaev model self-consistently and without assuming radiative efficiencies. The predicted theoretical sizes $r_{th}$ relative to a reference size $r_0$ (which is set to $r_0 =0$ lt-days) at a reference wavelength $\lambda_0$ can be written as: 
\begin{equation}
(r_{th}-r_0) \simeq 2 \ \left(\frac{L_{opt}}{10^{45}\,\mathrm{ergs~s^{-1}}}\right)^{1/2}\times\ \left[\left(\frac{\lambda}{\lambda_0}\right)^{4/3}-1\right]\, \mathrm{\, lt-days}.
\label{final_eq} 
\end{equation}

For a Shakura-Sunyaev profile, the characteristic size $r_s$ is related to the half-light radius by $R_{1/2} = 2.44\, r_s$. \citet{Ross2009} estimated a magnification-corrected luminosity at rest-frame at 1350\AA\ of $2.0\times10^{45}$ erg s$^{-1}$ based on power-law fits to the $B$, $V$, and $I$ HST images. \citet{Popovic2020} measured the fluxes of all four components from observations performed in 2018 (epoch 20) and obtained an average non-lensed quasar luminosity of $\lambda L(1350\AA) = (6.9 \pm 0.9) \times 10^{44}$ erg s$^{-1}$. By inserting the latest value for $L_{opt}$ in Eq. \ref{final_eq}, and adopting a reference wavelength of $\lambda_0 = 500$\AA\ (extreme UV; corresponding to the inner edge of the accretion disk), we find that the microlensing-based sizes are larger by a factor of $\sim2.5-4.3$ than the theoretical size estimates (see Table \ref{continuumsize}). Even after adding an external UV/X-ray term to Eq. \ref{final_eq} (as in \citealt{Fausnaugh2016}), assuming a local ratio of external to internal heating of 1 (i.e., the X-rays and viscous heating contribute equal amounts of energy to the disk), the sizes are only $\sim10\%$ larger, still unable to explain the discrepancy between the observed and the theoretical accretion disk sizes. This result is consistent with previous works, which have reported lensed quasar continuum emission regions larger than predicted by standard accretion disk theory (\citealt{Morgan2010,Blackburne2011,JimenezVicente2014}). This is also similar to the findings of optical continuum reverberation mapping campaigns of low-luminosity active galactic nuclei, which typically find that continuum emission region sizes are $\sim 2-3$ times larger than expected from disk reprocessing models (\citealt{Cackett2022}). One possible explanation for these larger-than-expected continuum sizes is a non-negligible contribution of diffuse continuum emission from the BLR to the observed continuum signals (e.g., \citealt{Cackett2018,Chelouche2019,Korista2019,Netzer2022}). It should be noted that our assumption is based on the signal ($L_{opt}$) lying within the wavelength range emitted by the self-similar parts of the disk. Although this assumption is justified for the rest-optical, it may not hold true for the rest-UV, particularly in cases where the SMBH mass is large or the mass accretion rate is small. In such scenarios, the turnover at shorter wavelengths, caused by the inner disk edge, shifts towards longer wavelengths. Consequently, our estimations of $r_{th}-r_0$ might only represent lower limits, implying that our disk over-estimations are conservative. 

\subsubsection{Size-wavelength relation}
\begin{figure}[h!]
\centering
\includegraphics[width=0.483\textwidth]{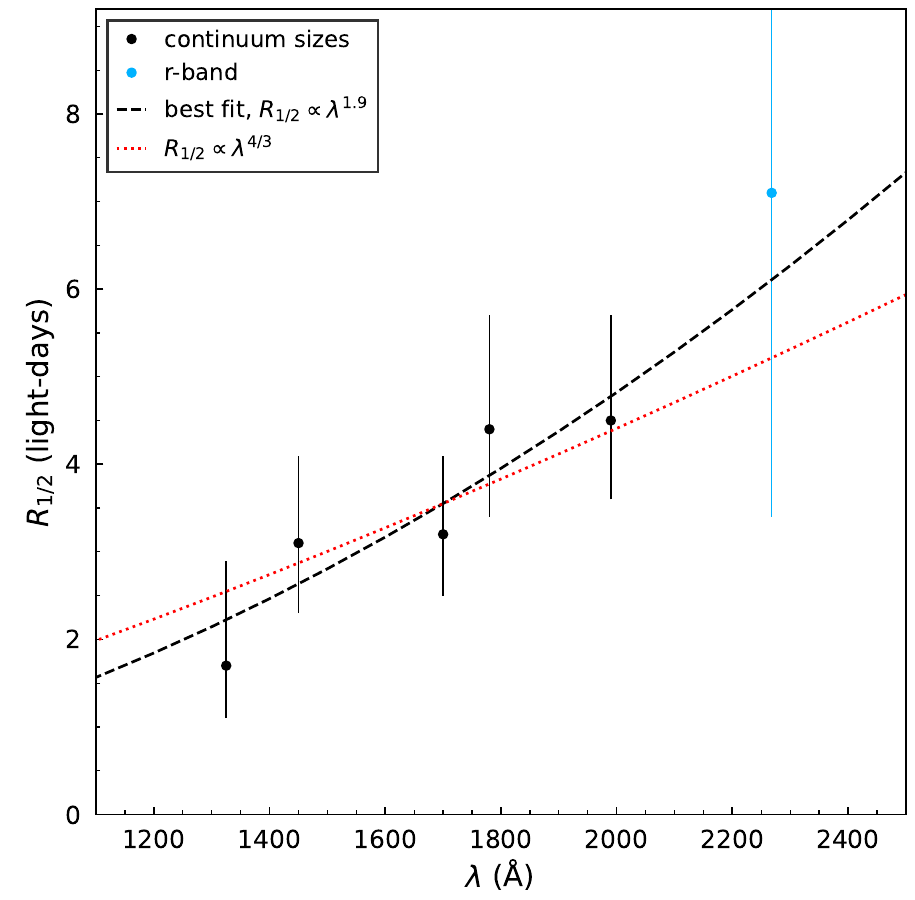}
\caption{Microlensing-based continuum-emitting sizes as a function of wavelength. The dashed black line shows the best fit to the data, with a power-law index of $\beta \sim 1.9$. The red dotted line is a fit with a fixed theoretical power-law index of $\beta = 4/3$, as expected for an optically thick and geometrically thin accretion disk.}
\label{r_lambda}
\end{figure}

Figure \ref{r_lambda} displays the microlensing-based continuum sizes as a function of wavelength. By fitting the estimated continuum-emitting sizes at different wavelengths with a disk model, we can infer the accretion disk size at a given wavelength (see Table \ref{continuumsize}). We fit our size spectrum with the power-law index $\beta$ (which quantifies the temperature profile of the disk) as a free parameter, obtaining the best fit with $\beta \sim 1.9$. From Figure \ref{r_lambda} we can see that the estimated sizes, as well as the physical model, are roughly consistent with the slope predicted for an optically thick and geometrically thin accretion disk model ($\beta = 4/3$). While several microlensing campaigns have found significantly larger sizes than predicted by the prevailing accretion disk theory (in agreement with the findings of this work), they frequently report flatter size-wavelength relations (\citealt{Morgan2010,Blackburne2011,JimenezVicente2014,Munoz2016}).

\begin{figure*}[h!]
\includegraphics[width=0.98\textwidth]{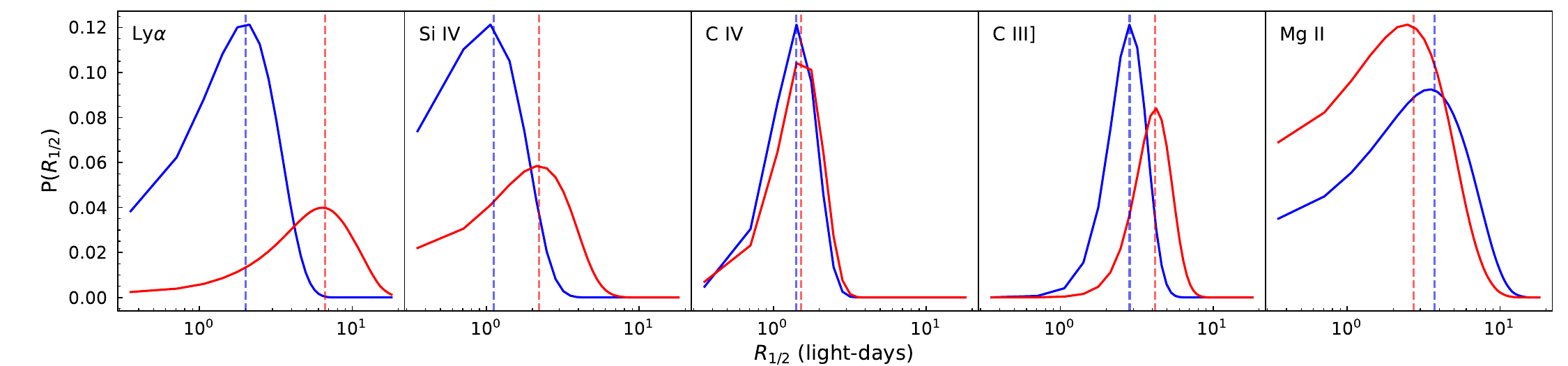}
\hspace*{2mm}
\caption{PDFs of the half-light radius $R_{1/2}$ emitting the blue and red wings of Ly$\alpha$, Si IV, C IV, C III], and Mg II. The vertical dashed lines indicate the expected size of the emission region using a uniform logarithmic prior. The emitting regions of high-ionization lines are smaller compared to those of low-ionization lines.}
\label{BEL}
\end{figure*}

\subsection{BLR size} 
We repeat the same procedures described in Section \ref{calcont} to infer the size of the continuum-emitting regions, but this time focusing on the wings of the BELs. In Figure \ref{BEL}, we present the PDFs corresponding to the regions emitting the broad wings of Ly$\alpha$, Si IV, C IV, C III], and Mg II. In Table \ref{BELsize}, we list the (minimal) size estimates along with their 68\% confidence intervals obtained for each of the BELs. Interestingly, we derive very small sizes for the regions emitting the BELs, sometimes even smaller than the sizes obtained for the region emitting the optical continuum. These findings are in disagreement (by an order of magnitude) with the BLR size estimates by \citet{Guerras2013}, as well as with the average BLR size obtained for a sample of lensed quasars in previous works (\citealt{Fian2018blr,Fian2021blr}). However, our estimated half-light radius of the C IV BLR aligns with the recent measurement reported by \citet{Hutsemekers2023} ($R_{1/2} = 2.8_{-1.7}^{+2.0}$ lt-days), indicating that our results are reliable. Typically, the BELs are expected to be less affected by microlensing than the continuum as they are emitted from a more extended region that is located further away from the central SMBH. One possible explanation for the uncommon observations could be a specific location on the magnification map. If the BLR, or part of the BLR, is located on or close to a caustic and the accretion disk is located away from it, the BLR will be highly magnified, while the accretion disk will be less magnified. This scenario is expected to be rare, as it depends on a very specific position and trajectory of the extended source on the caustic pattern, but could in principle explain the microlensing anomalies detected in this lensed system. An in-depth analysis of the BLR model, including the examination of specific source trajectories through the magnification pattern and their effect on microlensing of BEL wings and continuum-emitting sources, is beyond the scope of this current study and will be addressed in a future work.

\begin{table}[h]
	\tabcolsep=-0.005cm
	\renewcommand{\arraystretch}{1.4}
	\caption{Half-light radius $R_{1/2}$ of the BEL emitting regions.}
	\begin{tabu} to 0.49\textwidth {X[c]X[c]X[c]} 
		\hline
		\hline 
		Line & Feature & $R_{1/2}$ (lt-days) \\ 
		(1) & (2) & (3) \\ \hline
		Ly$\alpha$ & blue wing & $2.0_{-1.1}^{+1.4}$ \\
		            & red wing & $6.6_{-2.5}^{+4.7}$ \\ \hline
		  Si IV & blue wing & $1.1_{-0.4}^{+0.9}$ \\
		       & red wing & $2.2_{-1.0}^{+1.7}$ \\ \hline
        C IV & blue wing & $1.4_{-0.3}^{+0.6}$ \\ 
             & red wing & $1.5_{-0.4}^{+0.6}$ \\ \hline
        C III] & blue wing & $2.8_{-0.7}^{+1.0}$ \\ 
               & red wing & $4.2_{-1.0}^{+1.3}$ \\ \hline
        Mg II & blue wing & $3.7_{-1.7}^{+3.4}$ \\ 
              & red wing & $2.7_{-1.3}^{+2.7}$ \\ \hline
	\end{tabu}\\
	
	\small \textbf{Notes.} Cols. (1)--(2): BEL and emission feature. Col. (3): Half-light radius calculated using a logarithmic prior on the size, expressed in $\sqrt{M/0.3 M_\odot}$ lt-days.
\label{BELsize}	
\end{table}

\section{Conclusions}\label{6}
SDSS J1004+4112 is one of the most well-studied lensed quasars, with a wealth of photometric monitoring data and spectroscopic observations available. Despite its early identification as a lensed quasar displaying BEL deformation, a comprehensive understanding of the observed line distortions and differences between the lensed components remained elusive. In this work, we have made use of a recently published macro-model of the lens system (\citealt{Fores-Toribio2022}) and advanced techniques to quantitatively model the statistics of microlensing (see \citealt{Jimenez2022}). By compiling a sample of 20 spectroscopic observations, we have conducted a detailed analysis of the Ly$\alpha$, Si IV, C IV, C III], and Mg II lines, as well as their adjacent continua. The properties of BELs and their underlying continua provide crucial information on the nature of the BLR and accretion disk. Our results reveal various signatures of microlensing in the wings of the BELs, and through measurement of their strength, we can constrain the sizes of their emitting regions. The main conclusions of our study are the following: 
\begin{itemize}
\item[1.] This work has revealed consistency between the estimates of line core ratios between the lensed quasar images, which were used as a baseline for no microlensing, and the mid-IR and radio ratios reported in the literature (\citealt{Ross2009,Jackson2020,Hartley2021,McKean2021}).
\item[2.] We have found chromatic changes in the continuum adjacent to the BELs, providing insight into the structure of the accretion disk. The inferred continuum-emitting region sizes increase with wavelength, supporting the idea of disk reprocessing. The trend of size versus wavelength agrees with the prediction of a standard geometrically thin disk to some extent, however, the derived continuum-emitting sizes are larger than predicted by the Shakura-Sunyaev accretion disk model by a factor of $\sim 3.6$ on average. This discrepancy is consistent with recent findings (\citealt{JimenezVicente2014,Motta2017,Fian2016,Fian2018ad,Fian2021ad,Cornachione2020,Rojas2020}) and may be due to a substantial contribution of diffuse continuum emission from the BLR to the observed continuum signals (e.g., \citealt{Cackett2018,Chelouche2019,Korista2019,Netzer2022,Fian2023blr}). Furthermore, we want to emphasize that our inferred size for the region emitting the r-band continuum, as determined through spectroscopic data ($7.1_{-3.7}^{+7.4}$ lt-days), is in excellent agreement with the size inferred from 14.5 years of photometric monitoring data, with a half-light radius of $5.3_{-0.7}^{+1.3}$ lt-days (see also \citealt{Foresr2023submitted}).
\item[3.]Through a Bayesian analysis, we have determined the lower limits to the overall sizes of the regions emitting the BELs Ly$\alpha$, Si IV, C IV, C III], and Mg II. Our results indicate that the minimal sizes of these regions are on the order of a few light-days, which is notably smaller compared to the BEL emitting regions of an average quasar reported in the literature, as documented in previous studies (e.g., \citealt{Guerras2013,Fian2018blr,Fian2021blr}). These studies typically reveal sizes in the range of tens of light-days. The \mbox{C IV} BLR half-light radius estimated in our study is in line with that reported by \citet{Hutsemekers2023} ($R_{1/2} = 2.8_{-1.7}^{+2.0}$ lt-days), corroborating the validity of our measurements. Interestingly, for some of the emission lines, the inferred sizes are even smaller than the optical continuum-emitting size. One possible explanation for these unexpected results is that the BEL emitting regions are located near a caustic, while the accretion disk is farther away from it. Recently, \citet{Hutsemekers2023} demonstrated that the observed magnification profile of the C IV emission line in SDSS J1004+4112 can be reproduced using simple BLR models (i.e., a Keplerian disk or an equatorial wind). This suggests that the specific position and trajectory of the source through the magnification map plays a crucial role in determining the microlensing-based sizes of emitting regions in this system.
\end{itemize}

\begin{acknowledgements}
This research was supported by the grants PID2020-118687GB-C31, PID2020-118687GB-C32, and PID2020-118687GB-C33, financed by the Spanish Ministerio de Ciencia e Innovación through MCIN/AEI/10.13039/501100011033. J.A.M. is also supported by the Generalitat Valenciana with the project of excellence Prometeo/2020/085. J.J.V. is also supported by projects FQM-108, P20\_00334, and A-FQM-510-UGR20/FEDER, financed by Junta de Andalucía. D.C. and S.K. are financially supported by the DFG grant HA3555-14/1 to University of Haifa and Tel Aviv University, and by the Israeli Science Foundation grant no. 2398/19.
\end{acknowledgements}
\bibliographystyle{aa}
\bibliography{bibliography}

\end{document}